\documentclass[twocolumn,showpacs,aps,prd,amssymb,nofootinbib,nobibnotes,floatfix,longbibliography]{aastex631}

\usepackage{newtxtext,newtxmath}
\usepackage{ae,aecompl}
\usepackage[mathscr]{euscript}

\usepackage[T1]{fontenc}
\usepackage{hyperref}
\usepackage{amsfonts}
\usepackage{amsmath}
\usepackage{mathrsfs}
\usepackage{epsfig}
\usepackage{graphicx}
\usepackage{url}
\usepackage{hyperref}
\usepackage{float}
\usepackage{color}
\usepackage[utf8]{inputenc} 
\usepackage{epsf}
\usepackage{comment}
\usepackage{slashed}
\usepackage{footmisc}
\usepackage{lipsum}
\definecolor{linkcolor}{rgb}{0.0,0.3,0.5}
\usepackage[caption=false]{subfig}

\usepackage{color}
\definecolor{cerulean}{rgb}{0.0, 0.48, 0.65}

\definecolor{navy}{rgb}{0.2, 0.0, 1.0}

\definecolor{jungle}{rgb}{0.0, 0.5, 0.0}

\usepackage{color}

\definecolor{orange}{rgb}{1,0.5,0}
\definecolor{orangeB}{rgb}{1,0.7,0}

\usepackage[mathlines]{lineno}

\begin{document}

\preprint{APS/123-QED}

\title{Constraining Proper Motion of Strongly Lensed Eccentric Binary Mergers \\ using Doppler Triangulation}

\author{Johan Samsing}
\affiliation{Niels Bohr International Academy, The Niels Bohr Institute, Blegdamsvej 17, DK-2100, Copenhagen, Denmark}

\author{Lorenz Zwick}
\affiliation{Niels Bohr International Academy, The Niels Bohr Institute, Blegdamsvej 17, DK-2100, Copenhagen, Denmark}

\author{Pankaj Saini}
\affiliation{Niels Bohr International Academy, The Niels Bohr Institute, Blegdamsvej 17, DK-2100, Copenhagen, Denmark}

\author{Kai Hendriks}
\affiliation{Niels Bohr International Academy, The Niels Bohr Institute, Blegdamsvej 17, DK-2100, Copenhagen, Denmark}

\author{Rico K.~L.~Lo}
\affiliation{Niels Bohr International Academy, The Niels Bohr Institute, Blegdamsvej 17, DK-2100, Copenhagen, Denmark}

\author{Luka Vujeva}
\affiliation{Niels Bohr International Academy, The Niels Bohr Institute, Blegdamsvej 17, DK-2100, Copenhagen, Denmark}

\author{Georgi D. Radev}
\affiliation{The Niels Bohr Institute, Blegdamsvej 17, DK-2100, Copenhagen, Denmark}

\author{Yan Yu}
\affiliation{The Niels Bohr Institute, Blegdamsvej 17, DK-2100, Copenhagen, Denmark}

\shorttitle{Proper Motion of Strongly Lensed Eccentric Binary Mergers}
\shortauthors{Samsing et al.}

\date{\today}

\begin{abstract}

Strong lensing of gravitational wave (GW) sources allows the observer to see the GW source from different lines-of-sight (LOS)
through the corresponding images, which provides a way for constraining the relative proper motion of
the GW source. This is possible as the GW signals received from each image will have slightly different projected velocity components,
from which one can `Doppler-Triangulate' for the GW source velocity vector.
The difference in projected velocity between the different images can be observationally inferred through pairwise
GW phase measurements that accumulate over the time-of-observation. In this paper we study lensed eccentric GW sources and explore how the observable
GW phase shift between images evolve as a function of time, eccentricity, lens- and binary parameters. Next generation
GW observatories, including the Einstein Telescope and Cosmic Explorer, will see $\sim $hundreds/year of lensed GW sources,
where a significant fraction of these are expected to be eccentric. We discuss the expected unique observables for such
eccentric lensed GW sources, and the relation to their observable relative linear motion, which otherwise is exceedingly
difficult to constrain in general.

\end{abstract}

\section{Introduction}\label{sec:Introduction}

Observations of strong gravitationally lensed gravitational wave (GW) sources, provide not only information
about the lens, but also unique information about the GW source itself. One such unique
measure relates to the transverse proper motion of the GW source relative to the lens and the
observer \citep[e.g.][]{2009PhRvD..80d4009I, 2024PhRvD.109b4064S, 2024arXiv241214159S},
which can be constrained when two or more images are observed.
Recently in \cite{2024arXiv241214159S}, this was shown to be possible using ground-based detectors,
such as Einstein Telescope (ET) and Cosmic Explorer (CE), that are expected
to observe $\sim$hundreds of strongly lensed GW source per year \citep[e.g.][]{2022ApJ...929....9X, 2023MNRAS.520..702S}. Constraining the proper
motion of GW sources is exceedingly hard for GW mergers in general due to the absence of characteristic length- or mass scales associated
with individual events, as well as due to the difficulty in disentangling proper motion from the Hubble flow.
Observations of strongly lensed GW sources therefore have the potential to provide key insights into
aspects related to GW sources, their host environment, dynamics and possible
underlying formation mechanisms, which otherwise are not generally accessible through standard
single GW events.
    
\begin{figure}
    \centering
    \includegraphics[width=0.47\textwidth]{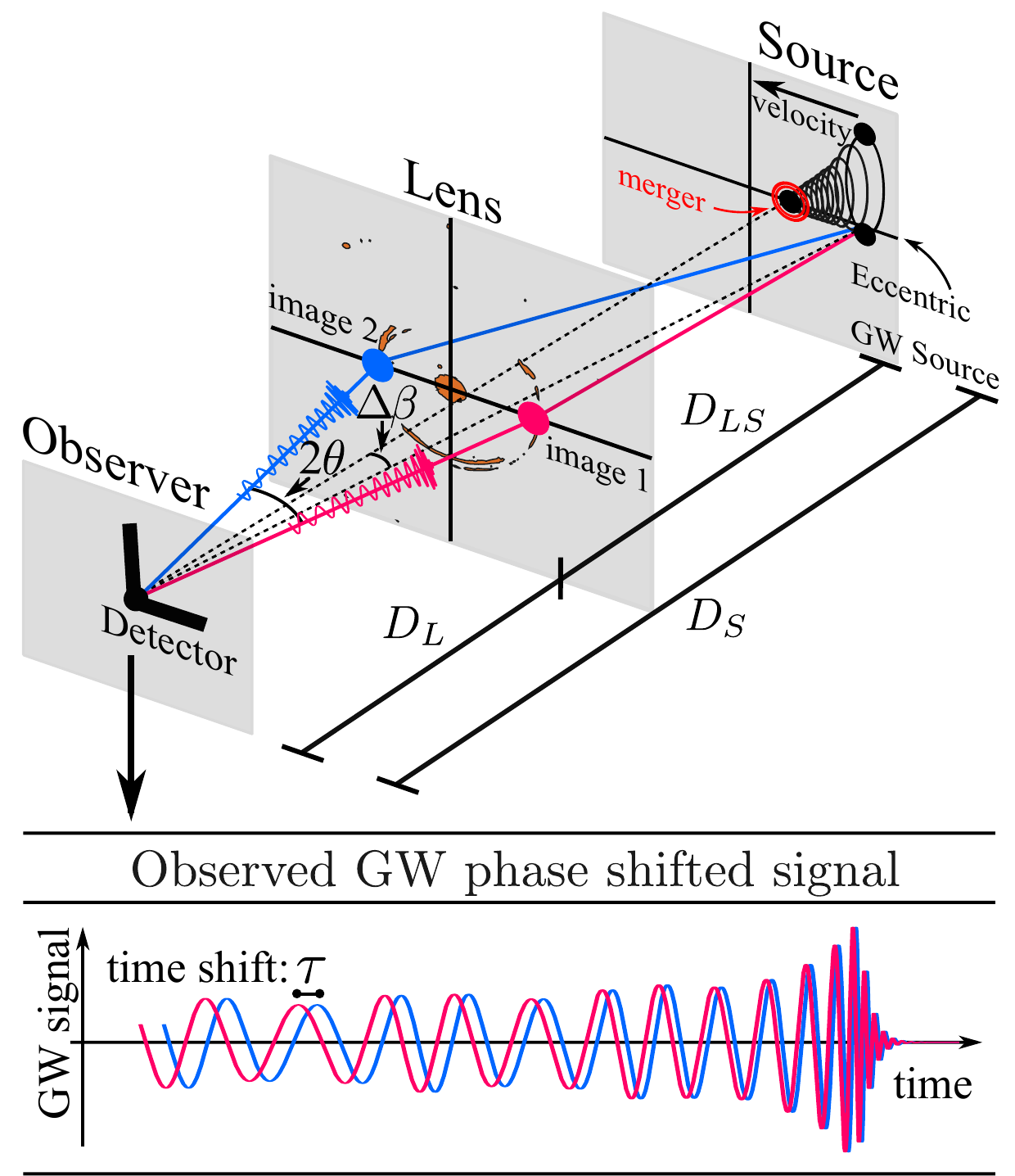}
    \caption{{\bf Illustration of a Strongly Lensed Eccentric Gravitational Wave Source.}
    The {\it Top Panel} shows the Observer-Lens-Source setup, with the detector in the {\it Observer}-plane, the
    gravitational lens in the {\it Lens}-plane, and the lensed eccentric GW source in the {\it Source}-plane. The
    eccentric GW source moves here in the source plane over a duration time $t$, that corresponds to a change in
    angular position $\Delta{\beta}$ and time-delay, $\tau$.
    The observer sees the lensed GW source through the two images; {\it image 1} and {\it image 2},
    that provide two different LOS towards the source, implying that the received signals generally will have slightly different
    Doppler factors. For GW sources, this manifests as a GW phase shift, as illustrated in the {\it Bottom Panel}, that can be observed and
    then related to the linear motion of the GW source relative to the lens and observer. If three of more images are observed, one can
    make a {\it Doppler-Triangulation} for better constraining the magnitude and direction of the velocity vector.
    }
    \label{fig:ill_lensingBBH}
\end{figure}

The relative transverse velocity of GW sources can be measured in strongly lensed events, as
each lens image essentially makes it possible for the observer to see the GW source from different
directional lines-of-sight (LOS) \citep{2024arXiv241214159S}.
As the velocity vector of the GW source projects differently onto the different LOS, the corresponding
GW images will have different Doppler factors, which implies that the received GW signals will
appear red- or blue-shifted relative to each other. In a flat static universe, the projected
difference in radial velocity as observed between two images, is by geometry (see Fig. \ref{fig:ill_lensingBBH})
given by $\Delta{v} \approx 2 \theta v$, where $v$ is the relative transverse velocity of the GW source,
and $2 \theta$ is the angular separation between the two lensed images.
This velocity difference gives rise to the relative Doppler factor, $\sim \Delta{v}/c$,
that can be linked to a displacement in angular phase of the received GW signals; a measurement
of GW phase shift between images can therefore be used to constrain the relative transverse motion of the GW source.
When three or more images are seen, one can triangulate for a better estimate of the direction of the GW source velocity
vector, explaining why we refer to this method as {\it Doppler-Triangulation}.
Variations of this method have been discussed a few places, both in terms of electromagnetic signals
\citep[][]{1986A&A...166...36K, 1989Natur.341...38C, 1989LNP...330...59B, 2004PhRvD..69f3001W}, as well as for
GW systems \citep[e.g.][]{2009PhRvD..80d4009I, 2020PhRvD.101h3031D,
2022MNRAS.515.3299G, 2024arXiv241016378Y, 2024PhRvD.109b4064S} with implications 
for deci-hertz detectors such as DECIGO/TianQin/Taiji \citep{2011CQGra..28i4011K, 2016CQGra..33c5010L, 10.1093/nsr/nwx116, 2020PhRvD.101j3027L},
and then recently with applications for ground-based detectors \citep{2024arXiv241214159S}.

Past studies and results have all been based on lensed circular GW sources. However, recent theoretical
advances in black hole dynamics hint that a significant fraction of GW sources likely evolve into the observable
bands with significant eccentricity \citep[e.g.][]{2006ApJ...640..156G, 2014ApJ...784...71S, 2017ApJ...840L..14S, Samsing18a, Samsing2018, Samsing18, 2018ApJ...855..124S, 2018MNRAS.tmp.2223S, 2018PhRvD..98l3005R, 2019ApJ...881...41L,2019ApJ...871...91Z, 2019PhRvD.100d3010S, 2020PhRvD.101l3010S, 2021ApJ...921L..43Z, 2022Natur.603..237S, Fabj24}.
As demonstrated in e.g. \cite{2024arXiv240305625S, 2024arXiv240804603H, 2024arXiv241108572H}, the GW phase
shift for accelerated eccentric GW sources, i.e. for sources with a $|dv/dt|>0$,
generally evolves very differently compared to circular GW sources. It is therefore of great importance to
understand how the GW phase shift between images of strongly lensed eccentric GW events evolve in our considered case of moving sources, and how the observables differ from the circular limit.

With this motivation, we here study for the first time the relationship between the relative proper motion of
strongly lensed eccentric GW sources and the resultant observable GW phase shift between the images.
Assuming a constant relative transverse motion between lens, GW source, and observer, we especially explore the evolution of the
GW phase shift as a function of time, GW frequency, and eccentricity. It is not yet understood if GW sources with different
formation pathways, e.g. dynamical formation relative to isolated binary formation, can be disentangled through their velocity flow linked to their host system.
However, our promising study serves as an excellent starting point for looking into this further, as well as 
to prepare for a future of learning from both eccentric- and circular strongly lensed GW sources.

The paper is organized as follows. In Sec. \ref{sec:Gravitational-Wave Phase shift} we describe the basics of inferring relative
transverse motion of strongly lensed GW sources from GW phase shift measurements between images. In Sec. \ref{sec:Eccentric Gravitational Wave Sources}
we provide our analytical solution applied to eccentric GW sources, after which we conclude our study in Sec. \ref{sec:Conclusions}.

\section{Constraining Proper Motion of Gravitational Wave Sources}\label{sec:Gravitational-Wave Phase shift}

\subsection{Theory and Observables}

We consider a setup composed of three planes; the {\it Observer plane}, the {\it Lens plane}, and
the {\it Source plane}, as shown in Fig. \ref{fig:ill_lensingBBH}. In an expanding universe, the relative velocities
translate non-trivially between these planes, as time and length scale with the expansion, or redshift, $z$.
Following \cite{1986A&A...166...36K, 2009PhRvD..80d4009I, 2024arXiv241214159S}, this implies that for the
general case where the observer, lens and source each move in their own respective frames
with transverse velocity $v_O$, $v_L$, and $v_S$, respectively, that one can define an effective velocity of the GW source
in its source plane as,
\begin{equation}
v' = v_S - \frac{1+z_S}{1+z_L}\frac{D_S}{D_L}v_L + \frac{1+z_S}{1+z_L}\frac{D_{LS}}{D_{L}}v_O,
\label{eq:v_prime}
\end{equation}
where $D$ refers to the angular diameter distance, and the subscripts `$O$', `$L$', and `$S$' refer to the
{\it O}bserver, {\it L}ens and {\it S}ource planes, respectively. This velocity $v'$ is the velocity
the GW source `appears' to have in the source plane, relative to a frame where the observer and lens are not moving. This also illustrates
that the GW phase shift we observe and here consider generally reflects information about the {\it combined} cosmologically
weighted relative velocities between observer, lens, and source. However, it is possible to subtract the observer
velocity, and if the lens is a massive galaxy cluster, then it can be argued that the velocity of the
GW source dominates.

For deriving the GW phase shift induced by the relative velocity $v'$, we now consider a stronly lensed GW source with two observed
GW images, or GW signals, that after observation have been aligned such that their time of merger coincides (shifted by the time-delay, $\Delta{t}$).
If the GW source has a non-zero velocity relative to the lens and observer, i.e. if $v' > 0$ from Eq. \ref{eq:v_prime},
then the two images will show a time-dependent temporal displacement from
the point-of-merger (see Fig. \ref{fig:ill_lensingBBH}). If we denote this time displacement $\tau$, then the corresponding
displacement in terms of orbital cycles in units of radians, referred to as the {\it GW phase shift},
can be expressed as,
\begin{equation}
\delta{\phi} = 2\pi{\tau}/{T}, 
\label{eq:dphi_def}
\end{equation}
where $T$ is the orbital time of the binary GW source \citep[e.g.][]{2024arXiv240305625S, 2024arXiv240804603H, 2024arXiv241108572H}.
Note here that $2/T$ is the GW frequency in the case of circular GW sources; however, this is not the case of eccentric
GW sources, as they emit with a much broader spectrum. The GW signal will however still be period with time $T$,
which is why we use this timescale in this general formulation of GW phase shift (see also \cite{2024arXiv240305625S}).

In our considered lensing case, the time $\tau$ is what encodes all the information about the lens setup and relative motion,
and can be estimated by considering the variation in the time-delay surface as the GW source is
moving. The kinematics of the movement relates to a change in the angular position of the GW source, $\Delta{\beta}$,
as follows (see Fig. \ref{fig:ill_lensingBBH}),
\begin{equation}
\Delta{\beta} = \frac{1}{1+z_S}\frac{v't}{D_S},
\end{equation}
where $t$ is defined in the observer frame, which then can be translated into $\tau$ as \citep{2024arXiv240305625S},
\begin{equation}
\tau \approx 2\theta \frac{D_L}{D_{LS}}\frac{1+z_L}{1+z_S}\frac{v'}{c}t,
\label{eq:tau}
\end{equation}
where $2\theta$ equals the angular separation between the two images, and $c$ is the
speed of light.
Substituting this into Eq. \ref{eq:dphi_def}, one now finds the general form for $\delta{\phi}$,
\begin{equation}
\delta{\phi} = 4 \pi {\theta} \frac{v_d}{c} \frac{t}{T(t)},
\label{eq:dphi_genform}
\end{equation}
where the velocity $v_d$,
\begin{equation}
v_d = v_O + \frac{D_L}{D_{LS}}\frac{1+z_L}{1+z_S}v_S - \frac{D_S}{D_{LS}}v_{L},
\label{eq:vd}
\end{equation}
represents an `effective Doppler velocity'.

In relation to the actual astrophysical observables, the GW phase shift in Eq. \ref{eq:dphi_genform} is the
key observable, but to constrain the value of $v_d$ one must also have an estimate for at least $\theta$. In electromagnetic observations of strong gravitational lensing, $\theta$ is
relatively easy to measure, but the extremely poor sky localization for GWs makes it near impossible for lensed GW sources. However, if a lens model is assumed,
$\theta$ can be inferred from other observables, namely the ratio between image magnifications and the time-delay between the lensed images \citep[e.g.][]{2009PhRvD..80d4009I}.
For example, for a Singular Isothermal Sphere (SIS) lens, the ratio between the magnification factors of the two observed images, $F_{12} = \mu_1/\mu_2 > 1$ can be written as,
\begin{equation}
F_{12} = \frac{\mu_1}{\mu_2} \approx \frac{\theta_E + \beta}{\theta_E - \beta},
\label{eq:F12}
\end{equation}
where $\theta_E$ is the Einstein angle (note that with our notation $2\theta_E \approx 2\theta$), and
$\beta$ is the angular position of the GW source. The corresponding time difference between the two
images, $\Delta{t}$, is given by
\begin{equation}
\Delta{t} \approx \frac{2D_LD_S}{D_{LS}}\frac{1+z_L}{c} \beta \theta_E.
\label{eq:Dt}
\end{equation}
By combining Eq. \ref{eq:F12} and Eq. \ref{eq:Dt}, one can now isolate for $\theta_E$,
\begin{equation}
\theta_E = \left( \frac{D_{LS}}{D_L D_S} \frac{c \Delta{t}}{2(1+z_L)} \frac{F_{12} + 1}{F_{12} - 1} \right)^{1/2},
\label{eq:thetaE}
\end{equation}
which then can be used in Eq. \ref{eq:dphi_genform} to approximate $\theta$.
For a better estimate one naturally has to employ a more sophisticated model for the
lens \cite[e.g.][]{2025arXiv250102096V} and an estimator for the angular diameter distances; however, this is beyond this work, and we therefore proceed with $\theta$ as our variable to keep our analysis more general.

\subsection{General Relations}

Assuming that $v_d$, and our lensing setup remain constant over the duration of observation, $t$,
it is clear from Eq. \ref{eq:dphi_genform} that the evolution of the GW phase shift
is dictated by how $T$ relates to $t$. As we are considering binary systems in this work, the time $T$ can be written as,
\begin{equation}
T(t) = 2\pi \sqrt{\frac{a(t)^3}{2Gm}}\,,
\label{eq:def_T}
\end{equation}
where $m$ is the mass of binary's components $(m_1 = m_2 = m)$, $a(t)$ is the semi-major axis (SMA) of the binary at time $t$. If we assume that
$a(t) \propto t^{\alpha}$, then the GW phase shift will follow the general scaling,
\begin{equation}
\delta{\phi} \propto t^{1-3\alpha/2}\ (a \propto t^{\alpha}).
\end{equation}
By now defining the critical value for $\alpha$, $\alpha_c = 2/3$, then for $\alpha > \alpha_c$ the GW phase shift $\delta{\phi}$
will {\it decrease} with time, whereas if $\alpha < \alpha_c$ then $\delta{\phi}$ will {\it increase}.
For example, for circular GW sources driven by GW radiation, $\alpha = 1/4$ \cite[e.g.][]{Peters64}, which leads
to $\delta{\phi} \propto t^{5/8}$. Depending on the (combination) of dissipative mechanisms for our considered lensed inspiraling
binary system, e.g. effects from gas or other external forces \citep[e.g.][]{2014barausse, 2023MNRAS.521.4645Z, 2024arXiv240305625S},
the evolution of the GW phase shift will show different characteristics. Below we will study properties of eccentric
evolving GW sources from formation to merger.

\begin{figure}
    \centering
    \includegraphics[width=0.49\textwidth]{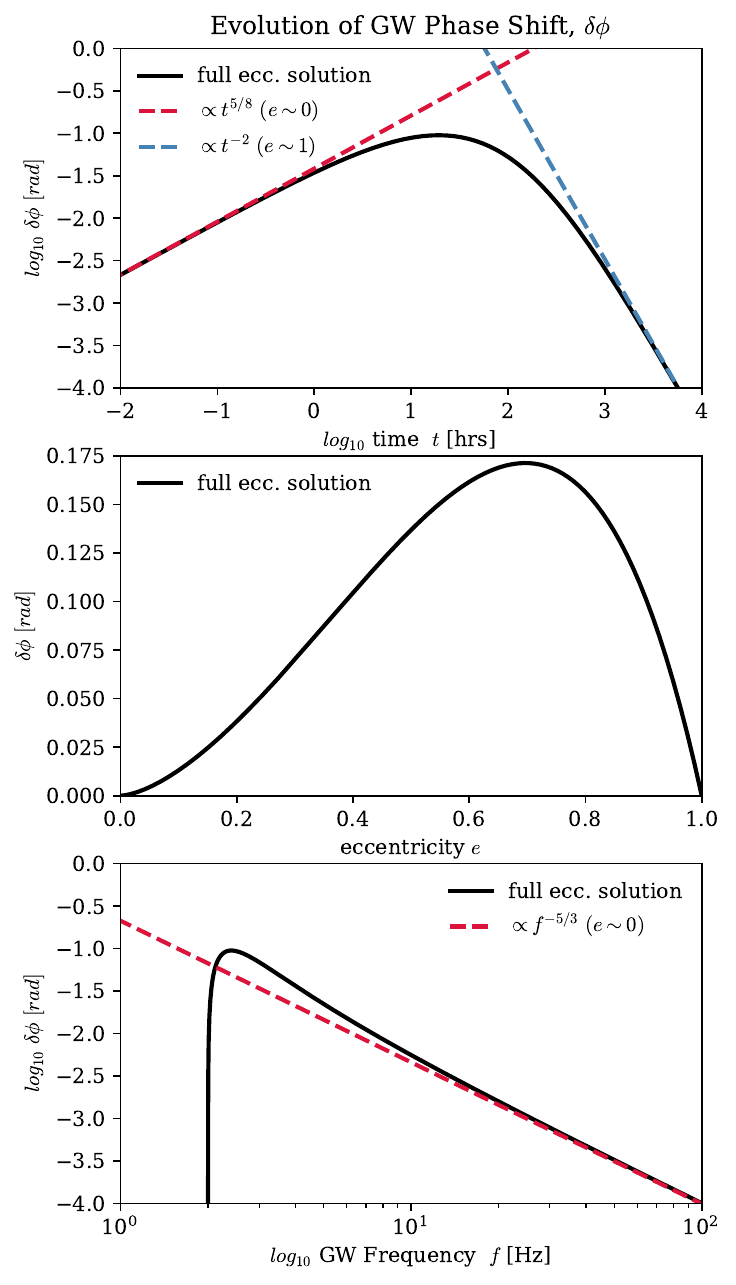}
    \caption{{\bf Evolution of Gravitational Wave Phase Shift:} The figure shows the GW phase shift, $\delta{\phi}$,
    as a function of time, $t$, eccentricity, $e$, and GW frequency, $f$, for our illustrative {\it Fiducial Model}
    outlined in Sec. \ref{sec:Eccentric Gravitational Wave Sources} ($v_d = 1500\ km/s$,
    $m = 5M_{\odot}$, $\theta = 25"$). Note here that $\delta{\phi} \propto \theta v_d$ in all our cases.
    {\it Top plot:} Here is shown $log \delta{\phi}, log t$ when going backwards from merger ($e \sim 0$) towards
    assembly ($e \sim 1$). The two asymptotic limits, $\delta{\phi} \propto t^{5/8} (e \sim 0)$ and $\delta{\phi} \propto t^{-2} (e \sim 1)$,
    are shown with {\it dashed lines}, where the {\it solid black line} is the result of solving \cite{Peters64} equations
    in combination with Eq. \ref{eq:dphi_ta32}. Note how the circular limit keeps growing, where the eccentric turns around.
    {\it Middle plot:} Evolution of $\delta{\phi}$ as a function of GW source eccentricity, $e$, given by Eq. \ref{eq:dphi_ecc_gen}.
    This suggests that the maximum GW phase shift will occur at $e \sim 0.7$.
    {\it Bottom plot:} plot of $log \delta{\phi}$ as a function of $log f$,
    where the {\it solid black line} is derived using \cite{Peters64} (note the approximate solution
    given by Eq. \ref{eq:dphi_f}), and the {\it red dashed line} shows the circular asymptotic limit, $ \delta{\phi} \propto f^{-5/3}$.
    Results are discussed in Sec. \ref{sec:Eccentric Gravitational Wave Sources}.    
    }
    \label{fig:dphi_tef}
\end{figure}

\section{Eccentric Gravitational Wave Sources}\label{sec:Eccentric Gravitational Wave Sources}

In the following we derive and study the evolution of the GW phase shift
induced by the relative transverse motion $v_d$, defined to Eq. \ref{eq:dphi_genform} and Eq. \ref{eq:vd}, for an eccentric GW source, as a function of time, eccentricity and GW frequency.
In all our derivations we assume for simplicity that each of the binary objects have equal and constant mass
$m$, from which it follows from Eq. \ref{eq:dphi_genform} and Eq. \ref{eq:def_T} that,
\begin{equation}
\delta{\phi} = \sqrt{8G} \times \theta \frac{v_d}{c} m^{1/2} \frac{t}{a(t)^{3/2}}.
\label{eq:dphi_ta32}
\end{equation}
The exercise therefore becomes to relate $a(t)$ and $t$ to the properties of the eccentric GW source, as we illustrate in the following.
In all our calculations we assume \cite{Peters64} relations for the evolution of the orbital elements, with a note on the
limits of this formalism \citep[e.g.][]{2020MNRAS.495.2321Z}.
For a more accurate calculation, one can e.g. use full GW forms and simply shift them by $\tau$ using Eq. \ref{eq:tau}.

For the examples shown in the following we also refer to our {\it Fiducial Model}, for which we assume the following
values for illustrative purposes; $v_d = 1500\ km/s$, $m = 5M_{\odot}$, $\theta = 25" (25\ arcsec)$, and an initial
GW peak frequency (see below) at formation of $f_0 = 2\ Hz$ assuming GW capture from a parabolic
orbit (see also \cite{2024arXiv240305625S})

\subsection{Evolution with Time}\label{sec:Evolution With Time}

We start by considering the GW phase shift as a function of time $t$, defined in the observer frame.
It is unfortunately not possible to algebraically isolate $\delta{\phi}$ as a function of $t$ using \cite{Peters64} for
the general eccentric case, but insight can be gained from considering the two limits for which the binary eccentricity approaches
$e = 0$ (near merger) and $e=1$ (near initial assembly). We do that in the following.

In the $e=0$ limit, the time $t$ relates to the binary parameters as \citep{Peters64},
\begin{equation}
t_c = \frac{5}{512} \frac{c^{5}}{G^{3}} \frac{a^4}{m^3},\ (e = 0).
\end{equation}
Now inserting this into the above Eq. \ref{eq:dphi_ta32}, one finds the GW phase shift to evolve as, 
\begin{equation}
\delta{\phi} = \sqrt{8} \left(\frac{5}{512}\right)^{3/8} \left(\frac{c^3}{G}\right)^{5/8} \theta \frac{v_d}{c} \frac{t^{5/8}}{m^{5/8}},\ (e = 0)
\end{equation}
which also was worked out in \cite{2024arXiv241214159S}. We now extend this to the eccentric limit,
where the time $t$ instead relates to $a(t)$ as \citep{Peters64},
\begin{equation}
t_e \approx t_c \times (1-e^2)^{7/2},\ (e > 0).
\label{eq:t_e}
\end{equation}
Note here that we have omitted the front factor $(768/425)$ to ensure our scalings naturally asymptote the circular limit 
when the GW source approaches merger.
As the binary peri-center distance, $r_p$, remains nearly constant in the eccentric limit, the merger time will here
instead scale as $t_e \propto r_p^{7/2} a(t)^{1/2} \propto a(t)^{1/2}$, where we have used $r_p = a(1-e)$. This
instead leads to the following relation,
\begin{equation}
\delta{\phi} \propto \theta \frac{v_d}{c} \frac{t^{-2}}{m^{5/8}},\ (e \sim 1).
\end{equation}
When going backwards in time from the point of merger, the GW phase shift therefore first increases $\propto t^{5/8}$,
until $\delta{\phi}$ reaches a maximum, after which it decreases $\propto t^{-2}$. These two scalings, together with
a curve derived by numerically evolving \cite{Peters64} coupled differential equations for $da/dt$ and $de/dt$,
are shown in the top plot of Fig. \ref{fig:dphi_tef} for
our {\it Fiducial Model} parameters.
The largest difference is clearly seen at earlier times, as the eccentric case decreases in contrast to the circular.
This essentially implies that steadily accumulating GW phase shift when going backwards in time from merger
is not always possible in the eccentric case, in contrast to the circular. We explore this further below.

\subsection{Evolution with Eccentricity}\label{sec:Evolution With Eccentricity}

The evolution of $\delta{\phi}$ can be written out analytically as a function of binary eccentricity, $e$,
which allows one to study properties such as maximum GW phase shift and eventually dependence on GW frequency.
For this we start by using the relation between $a$ and $e$ as given by \citep{Peters64},
\begin{equation}
a(e) \approx \frac{2r_0e^{12/19}}{(1-e^2)}\frac{g(e)}{g(1)},\ (e_0 \approx 1),
\label{eq:ae_e1lim}
\end{equation}
where 
\begin{equation}
g(e) = \left(1+121e^2/304\right)^{870/2299},
\label{eq:ge}
\end{equation}
and we have assumed that the initial eccentricity
$e_0 \approx 1$, and defined the initial peri-center distance $r_0  = a_0(1-e_0)$ \citep[e.g.][]{2024arXiv240305625S}.
By now using this relation together with Eq. \ref{eq:dphi_ta32} and Eq. \ref{eq:t_e}, we find that $\delta{\phi}$
can be expressed as a function of eccentricity as,
\begin{equation}
\delta{\phi} = \frac{2^{5/2}\sqrt{8}}{g(1)^{5/2}}\frac{5}{512} \frac{c^5}{G^{5/2}} \times \theta \frac{v_d}{c} \frac{r_0^{5/2}}{m^{5/2}} \times F(e),
\label{eq:dphi_ecc_gen}
\end{equation}
where the evolution is entirely encoded in the function,
\begin{equation}
F(e) = e^{30/19} (1-e^2) g(e)^{5/2}.
\label{eq:Fe}
\end{equation}
The maximum GW phase shift will therefore occur where $F(e)$ has its maximum, which is where the eccentricity $e$ equals,
\begin{equation}
e_m = \sqrt{\frac{\sqrt{107329}}{331} - \frac{167}{331}} \approx 0.7.
\label{eq:em}
\end{equation}
This is a moderate value, e.g. in the case of GW phase shift from accelerated sources $e_m \sim 0.95$ \citep{2024arXiv240305625S},
and should therefore be observable in the near future with the ongoing effort in modeling accurate eccentric GW forms.
The middle plot of Fig. \ref{fig:dphi_tef} shows $F(e)$ as a function of $e$.

\subsection{Evolution with Frequency}\label{sec:Evolution With Frequency}

\begin{figure}
    \centering
    \includegraphics[width=0.5\textwidth]{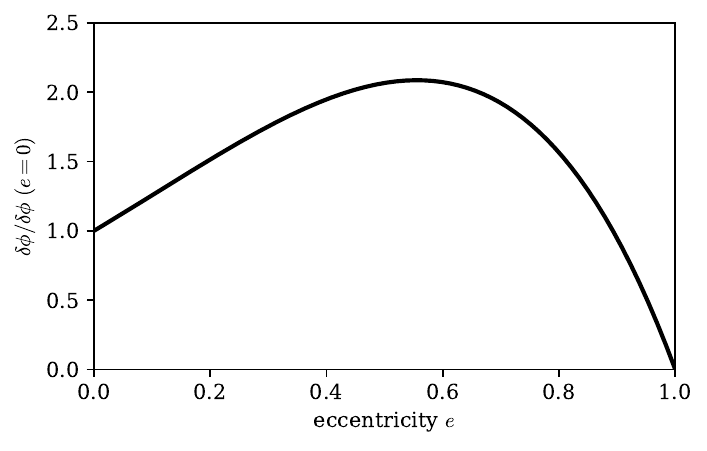}
    \caption{{\bf Eccentric GW Source Relative to Circular:} The figure shows the function $H(e)$ defined in Eq. \ref{eq:dphi_general} and
    Eq. \ref{eq:He}, which equals the ratio in GW phase shift between the eccentric case and the circular case, $\delta{\phi}/\delta{\phi}(e=0)$,
    evaluated at the same GW frequency $f$ (note here that $f$ in the eccentric case refers to the GW peak frequency defined in Eq. \ref{eq:fp}).
    As seen, when going backwards in time from merger $(e \sim 0)$, the ratio steadily rises until its reaches a maximum
    $\delta{\phi}/\delta{\phi}(e=0) \sim 2$ at $e \sim 0.6$, after which it decreases towards $0$ when $e \rightarrow 1$,
    or equivalently af $f=f_0$. The eccentric and circular cases can be seen separately in Fig. \ref{fig:dphi_tef}.
    }
    \label{fig:He}
\end{figure}

GW frequency for eccentric GW sources and how it relates to eccentricity and observables, is a matter of definition
\cite[e.g.][]{2024ApJ...969..132V}. Here we work with the GW frequency where most of the GW power is radiated over one orbit,
which is often referred to as the {\it GW peak frequency}, $f$. This can be approximated by,
\begin{equation}
f \approx \frac{1}{\pi} \sqrt{\frac{2Gm}{r_p^3}},
\label{eq:fp}
\end{equation}
where $r_p$ is the orbital peri-center distance. Note here that as $r_p$ stays approximately constant during inspiral, so will the GW peak frequency, until the binary has circularized and starts decaying as a standard circular source with $f \propto t^{-3/8}$.
With the above relation for $f$, we can now rewrite Eq. \ref{eq:dphi_ecc_gen} from above in the following forms,
\begin{align}
        \delta{\phi}    & = \frac{2^{10/3}\sqrt{8}}{g(1)^{5/2}{\pi^{5/3}}}\frac{5}{512}\frac{c^5}{G^{5/3}} \times \theta \frac{v_d}{c} {m^{-5/3}} {f_0^{-5/3}} \times F(e), \nonumber\\
                        & = \frac{2^{5/6}\sqrt{8}}{{\pi^{5/3}}}\frac{5}{512}\frac{c^5}{G^{5/3}} \times \theta \frac{v_d}{c} {m^{-5/3}} {f^{-5/3}} \times H(e), \nonumber\\
                        & = \delta{\phi}(e=0) \times H(e),
    \label{eq:dphi_general}
\end{align}
where
\begin{equation}
H(e) = (1+e)^{7/2}(1-e).
\label{eq:He}
\end{equation}
In the last equality of Eq. \ref{eq:dphi_general}, we have expressed the evolution in terms of the circular
limit $\delta{\phi}(e=0)$, which naturally is the asymptotic solution for $e \rightarrow 0$. From the above relations it follows
that $H(e)$ describes how much the eccentric case deviates from the circular limit. For example, the maximum value of
$\delta{\phi}$ relative to $\delta{\phi}(e=0)$ at a given GW frequency will be the maximum of $H(e)$,
\begin{equation}
max \left[\frac{\delta{\phi}}{\delta{\phi}(e=0)}\right] = \frac{10976\sqrt{14}}{19683} \approx 2,\ (e = \frac{5}{9} \approx 0.6),
\end{equation}
with corresponding value for $e$ shown in the parenthesis. The function $H(e)$ is shown in Fig. \ref{fig:He} as a function of $e$.
That the expected maximum increase in $\delta{\phi}$ is $\sim 2$ is not significant compared to e.g. a factor of $\sim 30$ for
accelerated sources \citep{2024arXiv240305625S}, but does illustrate that the eccentric limit at least is not suppressing the GW phase shift,
and should further have a unique shape that is observable.

For analytically being able to study how $\delta{\phi}$ evolves as a function of $f$ in our considered eccentric case,
we make the following approximation, $(f/f_0)^{-2/3} \approx e^{12/19}$, that leads to
$e \approx (f_0/f)^{19/18} \approx (f_0/f)$, which follows from Eq. \ref{eq:ae_e1lim} and valid in the
high eccentricity limit  \citep[e.g.][]{2024arXiv240305625S}. Substituting this into Eq. \ref{eq:dphi_general}
it now follows that,
\begin{equation}
\delta{\phi(f)} \approx \delta{\phi}(f,e=0) \times (1+f_0/f)^{7/2}(1-f_0/f).
\label{eq:dphi_f}
\end{equation}
This relation with the circular asymptotic limit is shown in the bottom plot of Fig. \ref{fig:dphi_tef}
for our {\it Fiducial Model} parameters. As seen, the eccentric limit rises above the circular limit as the eccentricity increases
towards lower $f$, as described by $H(e)$. When the eccentricity reaches the value $e_m$ given by Eq. \ref{eq:em}, the
GW phase shift rapidly starts decreasing for then to reach zero at $e \sim 1$, or equivalently when $f = f_0$. Depending on the
GW detector, only parts of this evolution might be observable, e.g. ET and CE would likely be able to observe most of the
shown evolution, whereas LIGO will see the signal only when $f > 10\ Hz$ at which the GW source has almost reached its circular asymptotic
limit and with a greatly reduced $\delta{\phi}$.

The GW frequency at which the maximum $\delta{\phi}$ is reached can be found by rewriting Eq. \ref{eq:ae_e1lim} in terms of $f$, such that 
\begin{align}
        \frac{f}{f_0}   & \approx \left( \frac{(1+e)}{2e^{12/19}}\frac{g(1)}{g(e)} \right)^{3/2}, \nonumber\\
                        & \approx 1.2,\ (e = e_m).  
    \label{eq:f_f0}
\end{align}
As seen, the GW frequency at max($\delta{\phi}$) is only higher by a factor of $\sim 1.2$ compared to the GW frequency at formation, $f_0$,
which tells that to observe the GW source near its maximum $\delta{\phi}$, the detector also have to operate down to frequencies $\sim f_0$.
For GW sources formed through dynamics, a significant fraction will distribute with $f_0$
between $1-10\ Hz$ \citep[e.g.][]{2019ApJ...871...91Z, 2020PhRvD.101l3010S}, and a significant part of
the evolution of $\delta{\phi}$, especially near its maximum, is therefore expected to be observed in the coming years.

\section{Conclusions}\label{sec:Conclusions}

The (relative transverse) proper motion can normally not be measured for GW sources, but
it is possible to constrain for strongly lensed GW sources, as the different images
allow the observer to see the GW source from different LOS. This is done by comparing
the GW phase evolution between images, or lensed GW signals, from which one can
infer a relative GW shift that can be mapped to the motion of the GW source.

In this paper, we explored for the first time the expected evolution of this GW phase shift
for a simple two-image lens model, as a function of time, binary eccentricity, and
GW peak frequency. We especially find that the GW phase shift evolution as a function of
eccentricity, $e$, takes a unique form
$\propto e^{30/10}(1-e^2)g(e)^{5/2}$ (Eq. \ref{eq:Fe}), that only depends on eccentricity,
from which we derived that the maximum GW phase shift occurs at $e \sim 0.7$ with an
enhancement of $\sim 2$ compared to the circular case.

A significant fraction of the expected GW mergers and lensed GW sources that we will observe in the coming decades with LIGO, ET and CE \citep[e.g.][]{2022ApJ...929....9X, 2023MNRAS.520..702S}, will be
eccentric \citep[e.g.][]{Samsing18, 2018PhRvD..98l3005R}. Our present study clearly
suggests that the GW phase shift should be possible to observe also for this population (for SNR calculations in the circular case see \cite{2024arXiv241214159S}),
which opens up immense possibilities, and the natural question if the eccentric GW population (dominated by the dynamical channel) is expected to have a different distribution in relative proper motion across redshift compared to the more generic circular GW population (dominated likely by the isolated binary evolution channel). In other words, can our suggested measure of
relative velocity be used for probing the nature and origin of GW sources?
In upcoming papers we will explore these questions both in relation to GW populations and assembly theory, as well
as the possible impact that a measure of transverse velocity for hundreds of lensed GW sources could have on cosmology.

\section{Acknowledgments}

The authors are grateful Juan Urrutia, Mikołaj Korzyński, Miguel Zumalacárregui, and Graham Smith, for useful discussions.
We further thank the The Erwin Schrödinger International Institute for Mathematics and Physics (ESI)
and the organizers of the workshop "Lensing and Wave Optics in Strong Gravity" where part of this work was
carried out.
K.H, L.Z., P.S., and J.S. are supported by the Villum Fonden grant No. 29466, and by the ERC Starting
Grant no. 101043143 -- BlackHoleMergs led by J. Samsing.
R.L and L.V. are supported by the research grant no. VIL37766 and no. VIL53101 from Villum Fonden, and the DNRF Chair
program grant no. DNRF162 by the Danish National Research Foundation. 

\bibliographystyle{aasjournal}
\bibliography{NbodyTides_papers}

\end{document}